\documentclass[10pt]{iopart}
\usepackage{graphicx}
\usepackage{mathptmx}
\usepackage{color}
\usepackage{multicol}
\usepackage{cite}
\usepackage{amssymb}
\usepackage{algorithmicx}
\usepackage{algorithm}
\usepackage{algpseudocode}
\usepackage{subcaption}
\usepackage{calc} 

\usepackage{nameref,hyperref}
\usepackage[switch]{lineno}
\usepackage{microtype}
\DisableLigatures[f]{encoding = *, family = * }
\usepackage[table]{xcolor}
\usepackage{array}
\usepackage[inline]{enumitem}
\usepackage{cleveref}
\usepackage{tikz}
\usepackage{placeins}
\usepackage{float}
\usetikzlibrary{shapes,arrows,positioning}

\begin{document}

\title[$\mathrm{K~Vaziri}$ et al]{\huge{Nonlinear Dynamical Modeling of Human Intracranial Brain Activity with Flexible Inference}}

\author{Kiarash Vaziri\textsuperscript{1}, Lucine L. Oganesian\textsuperscript{1}, HyeongChan Jo\textsuperscript{1}, Roberto M. C. Vera\textsuperscript{2}, Charles Y. Liu\textsuperscript{2}, Brian Lee\textsuperscript{2}, and Maryam M. Shanechi\textsuperscript{1,3*}}

\address{\textbf{1} Ming Hsieh Department of Electrical and Computer Engineering, Viterbi School of Engineering, University of Southern California, Los Angeles, CA, USA}
\address{\textbf{2} Keck School of Medicine, University of Southern California, Los Angeles, CA, USA}
\address{\textbf{3} Departments of Computer Science and Biomedical Engineering, and the Neuroscience Graduate Program, University of Southern California, Los Angeles, CA, USA}
\address{ * Author to whom any correspondence should be addressed.}
\ead{shanechi@usc.edu}
\begin{abstract}
~\\
\small{
Dynamical modeling of multisite human intracranial neural recordings is essential for developing neurotechnologies such as brain-computer interfaces (BCIs). Linear dynamical models are widely used for this purpose due to their interpretability and their suitability for BCIs. In particular, these models enable flexible real-time inference, even in the presence of missing neural samples, which often occur in wireless BCIs. However, neural activity can exhibit nonlinear structure that is not captured by linear models. Furthermore, while recurrent neural network models can capture nonlinearity, their inference does not directly address handling missing observations. To address this gap, recent work introduced DFINE, a deep learning framework that integrates neural networks with linear state-space models to capture nonlinearities while enabling flexible inference. However, DFINE was developed for intracortical recordings that measure localized neuronal populations. Here we extend DFINE to modeling of multisite human intracranial electroencephalography (iEEG) recordings. We find that DFINE significantly outperforms linear state-space models (LSSMs) in forecasting future neural activity. Furthermore, DFINE matches or exceeds the accuracy of a gated recurrent unit (GRU) model in neural forecasting, indicating that a linear dynamical backbone, when paired and jointly trained with nonlinear neural networks, can effectively describe the dynamics of iEEG signals while also enabling flexible inference.  Additionally, DFINE handles missing observations more robustly than the baselines, demonstrating its flexible inference and utility for BCIs. Finally, DFINE's advantage over LSSM is more pronounced in high gamma spectral bands. Taken together, these findings highlight DFINE as a strong and flexible framework for modeling human iEEG dynamics, with potential applications in next-generation BCIs.
}
\end{abstract}

\ioptwocol
\section{Introduction}\label{sec: Intorduction}
Modeling neural population activity in human recordings is of broad interest across various neuroscienctific and neural engineering domains -- both for studying the brain and for developing neurotechnologies, such as brain-computer interfaces (BCIs) \cite{engel2005_invasive, brandmanReviewHumanIntracortical2017, shanechiBrainMachineInterfaces2019a, llo_reviewpaper_2024}. A prominent modality for directly measuring the human brain is intracranial electroencephalography (iEEG), which can capture multisite field potentials. 
Indeed, iEEG supports continuous recording from multiple brain regions simultaneously over long time periods\cite{karumbaiahRelationshipIntracorticalElectrode2013,changLargeScaleHumanBasedMesoscopic2015, woodsLongtermRecordingReliability2018}, thereby enabling the study of the network-level dynamics underlying complex behaviors and internal states such as mood \cite{sani2018_mood_decoding}, cognition \cite{lachaux2012_hfn_cognition, jacobsHippocampalThetaOscillations2014, johnsonIntracranialRecordingsHuman2015, basuClosedloopEnhancementNeural2023}, and speech \cite{anumanchipalli2019_speech_synthesis,bhaya-grossmanSharedLanguagespecificPhonological2025}, as well as disorders such as epilepsy \cite{jacobsHighfrequencyOscillationsHFOs2012, zhangLowComplexitySeizurePrediction2016}, chronic pain \cite{shirvalkarFirstinhumanPredictionChronic2023}, major depression \cite{shanechiBrainMachineInterfaces2019a, llo_reviewpaper_2024 ,sani2018_mood_decoding,scangosClosedloopNeuromodulationIndividual2021,alagapanCingulateDynamicsTrack2023}, post-traumatic stress disorder \cite{gill2023pilot}, eating disorder \cite{shivacharan2022pilot},  and obsessive-compulsive disorder \cite{mosesleeInvasiveBrainMapping2025,nho2024responsive,provenza2024disruption}. Moreover, chronic iEEG recordings enable long-term monitoring of brain dynamics. Such recordings can be useful for studying the neural mechanisms underlying prolonged treatments, such as deep brain stimulation \cite{saalIndividualizedDeepBrain2025}, as well as for characterizing disease states and symptoms across both short and long timescales \cite{ashkanInsightsMechanismsDeep2017,llo_reviewpaper_2024}. Taken together, intracranial neural recordings offer a unique window through which complex, multiregional brain dynamics underlying various behavioral and internal states can be studied. However, to enable such analyses, rich and robust models of multisite iEEG data are required. 

Dynamical system models, with their ability to simultaneously describe both the temporal evolution of high-dimensional neural activity and the spatial relationship among electrodes, offer a framework with which to model iEEG data. In particular, \emph{latent state dynamical models} can capture the spatiotemporal structure in neural activity by learning low-dimensional latent state variables that summarize past observations and evolve over time. Due to their dimensionality reduction and autoregressive properties, such dynamical models are of broad interest for both neuroscientific and neural engineering applications. For example, linear dynamical models have been developed to describe the spatiotemporal dynamics of iEEG \cite{yang_lssm_2019,ahmadipour2021_adaptive_tracking, nozari2024macroscopic}. Furthermore, they have been used to identify neural biomarkers of various affective states, such as mood \cite{sani2018_mood_decoding} and chronic pain \cite{shirvalkarFirstinhumanPredictionChronic2023} states. Moreover, latent state dynamical models have potential utility for designing closed-loop BCI and neurostimulation systems \cite{shanechiBrainMachineInterfaces2019a,llo_reviewpaper_2024,hoang2017biomarkers,yangControltheoreticSystemIdentification2018,yang2021modelling,bolusStatespaceOptimalFeedback2021}. 

However, in order to enable the design of robust BCI systems with iEEG measurements, a latent state dynamical model of iEEG should satisfy two properties. First, the model should achieve accurate nonlinear modeling of the underlying neural dynamics. Second, the model should allow for flexible inference, meaning it should enable real-time recursive inference even in the presence of missing neural samples, which can often happen, for example, in wireless BCIs. Despite the demonstrated progress in modeling iEEG data, this objective of simultaneously enabling nonlinear dynamical modeling and flexible inference has not yet been addressed when modeling multisite intracranial human neural recordings.

One body of prior work on latent state dynamical modeling of iEEG has focused on training linear state-space models (LSSMs) \cite{yang2019developing,ahmadipour2021_adaptive_tracking,sani2018_mood_decoding,shirvalkarFirstinhumanPredictionChronic2023, nozari2024macroscopic}. One major strength of these models is their ability for flexible inference with Kalman filtering \cite{Astrom2012StochasticControl}. Given this flexible inference property, as well as their interpretability, LSSMs have been widely applied in BCI systems \cite{Gilja2012HighPerformance, Sussillo_2012, Nason2021Realtime,kaoSingletrialDynamicsMotor2015,  shanechi2016robust,shanechi2017rapid, nason2020low, orsborn2012closed} and have been effectively used to model and predict intracranial human neural activity \cite{sani2018_mood_decoding,yang2019developing,ahmadipour2021_adaptive_tracking,shirvalkarFirstinhumanPredictionChronic2023, nozari2024macroscopic}. Despite these nice properties, LSSMs are limited for describing the nonlinearities present in neural data. Another class of latent state dynamical models that have been used with iEEG are recurrent neural networks (RNNs), which incorporate nonlinear dynamics in their autoregressive backbone \cite{anumanchipalli2019_speech_synthesis, duraivelHighresolutionNeuralRecordings2023, nozari2024macroscopic}. However, RNNs do not directly address inference with missing neural observations \cite{anumanchipalli2019_speech_synthesis, duraivelHighresolutionNeuralRecordings2023, nozari2024macroscopic}, which is especially important for building robust BCI systems.

Towards addressing this challenge of simultaneously modeling nonlinearity in neural data while enabling flexible inference  in the presence of missing samples, recent work introduced dynamical flexible inference for nonlinear embeddings (DFINE) \cite{dfine_2024}. DFINE combines the expressive power of neural networks for capturing nonlinearities, with a linear state-space model backbone for flexible inference. However, DFINE was developed for modeling the dynamics in localized neuronal populations that are measured with intracortical arrays in non-human primates \cite{dfine_2024}. As such, whether or not DFINE can be extended to model multisite human iEEG dynamics and how accurately remain unexplored. 

Thus, to address these questions, in this work we extend DFINE to modeling a new modality, multisite intracranial recordings from human subjects, and evaluate its performance across datasets from 10 individuals. We quantify the performance of DFINE and other benchmark latent state dynamical models of iEEG in terms of how well they forecast future iEEG activity from its past observations. As baseline models, we compare against LSSMs and RNNs given their broad utility in neuroscience and prior demonstrations on iEEG data. First, we show that DFINE provides a significant improvement in forecasting iEEG power features compared with the previously established LSSM baseline. This improvement demonstrates that DFINE can more accurately capture iEEG dynamics than LSSM, while preserving the key flexible inference advantage of LSSMs. Second, we find that DFINE performs on par with (better than or similar to) a nonlinear RNN for predicting iEEG power features. This result suggests that DFINE captures the nonlinear neural dynamics competitively with RNNs, while also enabling flexible inference -- which is not supported by RNNs. Third, we show that DFINE maintains enhanced accuracy across longer forecasting horizons and outperforms the baselines in doing so, underscoring the robustness of its latent dynamical backbone for long-term neural prediction. Fourth, we show that DFINE achieves superior neural prediction accuracy even when portions of the iEEG signal are missing, demonstrating its robustness to signal dropout, which can frequently arise in real-world wireless BCI use cases. Finally, we observe that DFINE provides its largest performance gains over LSSM when predicting iEEG power in the higher-frequency bands (65–100 Hz, high gamma) compared to the other lower frequency bands.

Together, these findings position DFINE as a powerful framework for modeling human intracranial activity, with broad implications for the development of translational BCI systems.

\section{Methods}\label{sec: Methods}
\subsection{Problem setup}\label{Methods: Dynamic models}
Let time-series $\{y_t \in \mathbb{R}^{n_y}, \; t=1, 2, ..., T\}$ denote the neural activity over time $T$, where $n_y$ is the number of observed signals. In our case, $y_t$ corresponds to spectral power features that are computed for multiple frequency bands per iEEG contact (Section~\ref{methods-powerfeatures}). Here we aim to learn a latent state dynamical model of the neural activity that can be used towards neural forecasting. Formally, such models are often expressed as  
\begin{equation} \label{eq:general-dynamics}
\left\{
\begin{array}{cc}
    x_{t+1} = g(x_t),\\
    y_t = h(x_t)
\end{array}
\right.
\end{equation}
where $x_t \in \mathbb{R}^{n_x}$ denotes the latent state, and $g(
\cdot)$ and $h(\cdot)$ define the state transition and observation mappings, respectively. Within this framework, forecasting amounts to predicting future observations based on the past. In the one-step-ahead case, the prediction of the next observation can be written as  
\begin{equation}
    \hat{y}_{t+1|t} = \mathcal{F}_1(y_1, y_2, \ldots, y_t),
\end{equation}
where $
\mathcal{F}(\cdot)$ denotes the prediction function. A multi-step-ahead predictor can be defined similarly as 
\begin{equation}
    \hat{y}_{t+K|t} = \mathcal{F}_K(y_1, y_2, \ldots, y_t),
\end{equation}
where $K$ is the number of steps into future (i.e., the forecasting horizon). 

To evaluate the predictive accuracy of a learned model, we use the \emph{normalized root mean squared error (NRMSE)}, defined as:
\begin{equation} \label{eq:nrmse}
    \mathrm{NRMSE} = 
    \frac{\sqrt{\sum_t (y_t - \hat{y}_{t})^2}}
         {\sqrt{\sum_t (y_t - \bar{y})^2}},
\end{equation}
where $y_t$ and $\hat{y}_t$ denote the observed and predicted signals at time $t$, and $\bar{y}$ is the mean of the observed signal across all time $T$. The normalization in the denominator rescales the error relative to the variance of the data, ensuring that performance is not dominated by the overall signal amplitude. As a result, NRMSE provides a dimensionless measure of accuracy that allows comparison across datasets with different signal scales. An $\mathrm{NRMSE}$ of $0$ indicates perfect prediction, while $\mathrm{NRMSE}$ of $1$ corresponds to predictions no better than a constant mean predictor. Values greater than $1$ reflect performance worse than the mean predictor.
In this work we use DFINE as a nonlinear and flexible dynamical model and compare its prediction performance with a naive predictor baseline and two dynamical models commonly used in neuroscience: a latent linear state-space model and a gated recurrent unit (Section \ref{methods-baselines}).

\subsection{DFINE}\label{Methods: DFINE}
\subsubsection{Model architecture}
DFINE is a neural network architecture designed to capture nonlinear structure in neural data while retaining flexible inference, similar in spirit to linear state-space models \cite{dfine_2024}. DFINE has two distinct sets of latent state variables: dynamic latent factors $x_t \in \mathbb{R}^{n_x}$ and manifold latent factors $a_t \in \mathbb{R}^{n_a}$. The dynamic factors evolve according to a linear dynamical system capturing smooth temporal dependencies, whereas the manifold factors provide a nonlinear embedding that mediates between the latent dynamics and the observed nonlinear neural activity, $y_t \in \mathbb{R}^{n_y}$. This separation enables DFINE to model complex, nonlinear structure in the data while maintaining flexible inference in the latent dynamical space.

Formally, the dynamic latent factors evolve according to a linear Gaussian state-space model,
\begin{equation} \label{methods-dfine-lssm-xt}
    x_{t+1}=Ax_t+w_t,
\end{equation}
where $w_t \in \mathbb{R}^{n_x}$ is zero-mean Gaussian noise with covariance $W\in \mathbb{R}^{n_x\times n_x}$, and $A\in \mathbb{R}^{n_x\times n_x}$ is the state transition matrix (e.g., $g(\cdot)$ in equation (\ref{eq:general-dynamics})). The latent manifold factors are then linked to the dynamic factors through a noisy linear mapping,
\begin{equation} \label{methods-dfine-lssm-at}
    a_t=Cx_t+v_t,
\end{equation}
with $C\in \mathbb{R}^{n_a \times n_x}$ representing the emission matrix and $v_t \in \mathbb{R}^{n_a}$ zero-mean Gaussian noise with covariance $R\in \mathbb{R}^{n_a\times n_a}$. Together, equations (\ref{methods-dfine-lssm-xt}) and (\ref{methods-dfine-lssm-at}) form an LSSM, where the manifold factors, $a_t$, act as noisy Gaussian observations of the underlying dynamics, $x_t$. The parameters of this LSSM are given by $\psi=\left\{A, W, C, R, \mu_0, \Lambda_0\right\}$, where $\mu_0$ and $\Lambda_0$ specify the mean and covariance, respectively, of the initial dynamic factor state distribution (i.e., $x_0 \sim \mathcal{N}(\mu_0, \, \Lambda_0)$). 

To capture nonlinear structure, DFINE employs an autoencoder that relates observations $y_t$ to the manifold latent factors $a_t$. The autoencoder consists of an encoder that maps neural observations to the manifold factors ($f_\phi:y\rightarrow a$) and a decoder that mirrors the structure of the encoder to reconstruct neural observations from the manifold factors ($f_{\theta}:a\rightarrow y$). 
The encoder and decoder are modeled as nonlinear functions, such that 
\begin{equation} \label{methods-dfine-encoder}
    a_t=f_{\phi}(y_t)
\end{equation}
\begin{equation} \label{methods-dfine-decoder}
    y_t=f_\theta(a_t)
\end{equation}
where $\phi$ and $\theta$ represents the encoder and decoder parameters, respectively. Here we take the encoder and decoder models to be multilayer perceptrons (MLPs) due to their universal function approximation capabilities \cite{hornik1989multilayer}.

In all analyses we set the manifold factor dimensionality to be $n_a = 200$, to match the number of input features $n_y$, and use a 1-layer MLP for the encoder and decoder (see Appendix Figure \ref{app:mlp-blocks})
We fix the dynamic factor dimensionality across all models to be $n_x = 32$, thus promoting learning of a low-dimensional latent model but without loss of neural prediction accuracy. (In Appendix \ref{app:nx}, we show that similar results hold across different values of $n_x$.) 

Taken together, equations (\ref{methods-dfine-lssm-xt})-(\ref{methods-dfine-decoder}) define the full structure of DFINE (also depicted in Figure \ref{fig:DFINE-architecture}a). It should be emphasized that all parameters of DFINE are trained jointly in an end-to-end manner (Section \ref{Methods: DFINE-parameter optimization}). Through this joint manifold-dynamics learning, the autoencoder is forced to learn the best nonlinear manifold embedding over which the latent dynamics can be described with a linear dynamical backbone.

\subsubsection{Future Prediction} 
\begin{figure*}
    \centering
    \includegraphics[width=1\linewidth]{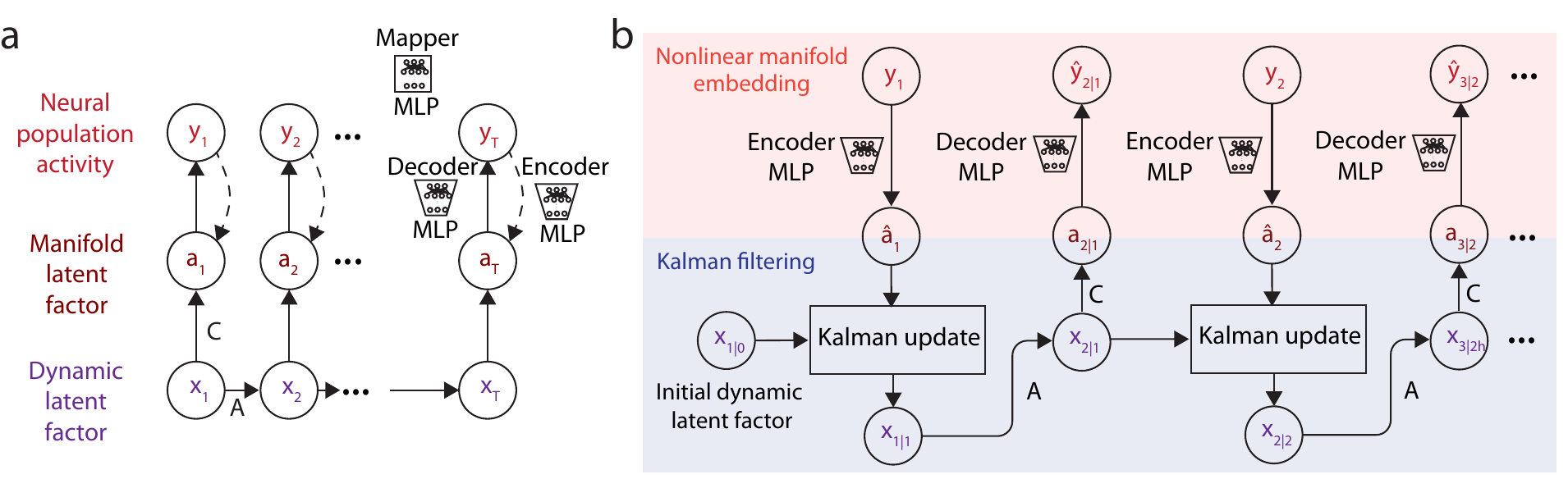}
    \caption{Model architecture of DFINE. (a) The generative model of DFINE is shown. DFINE uses two sets of latent factors: the dynamic and manifold latent factors. The dashed connection from neural observations to the manifold latent factor is used only during inference. (b) The inference procedure with DFINE is shown. We first estimate the manifold latent factors using the nonlinear manifold embedding at each timestep. Using the dynamic equation, we recursively run the Kalman filter to infer the dynamic latent factor $x_{t|t}$ and predict the next dynamic latent factor by using the Kalman predictor to obtain first $x_{t+1|t}$ and then $a_{t+1|t}$. Finally, to predict the observation in the future, DFINE maps Kalman predicted manifold factor through the decoder to estimate $\hat{y}_{t+1|t}$. All model parameters (LDM and autoencoder) are trained jointly to learn the best nonlinear embeddings over which the dynamics can be linear.}
    \label{fig:DFINE-architecture}
\end{figure*}

To perform future prediction, neural activity $y_t$ is first statically mapped to manifold factor $\hat a_t$ using the encoder in equation (\ref{methods-dfine-encoder}). Then, using $\hat a_t$ and the latent recurrence defined by equation (\ref{methods-dfine-lssm-xt}), the state of the dynamic factors at the next timestep, i.e., $\hat{x}_{t+1|t}$, is estimated using a Kalman predictor \cite{Astrom2012StochasticControl}. Finally, the one-step-ahead prediction of the manifold factor (i.e., $\hat{a}_{t+1|t}$) is computed using the observation mapping defined in equation (\ref{methods-dfine-lssm-at}). This entire process is formalized as
\begin{equation} \label{methods:kalman-prediction}
\left\{ 
\begin{array}{ccl}
    \hat x_{t+1|t}&=&A\hat x_{t|t-1}+G_{t}(y_{t}-C\hat x_{t|t-1})\\
     \hat{a}_{t+1|t}&=&C\hat{x}_{t+1|t}\\
\end{array}
\right.
\end{equation}
where $G_{t}\in \mathbb{R}^{n_x\times n_y}$ is the Kalman gain defined as
\begin{equation}
    G_{t}=AP_{t|t-1}C^T(CP_{t|t-1}C^T+R)^{-1}.
\end{equation}
and $P_{t|t-1}$ is the prediction error covariance \cite{Astrom2012StochasticControl}, computed recursively as
\begin{equation}
\left\{
\begin{array}{ccl}
P_{t|t-1} &=& A P_{t-1|t-2} A^{T} \\[6pt]
&&\displaystyle 
-\, A P_{t-1|t-2} C^{T} \\[4pt]
&&\displaystyle 
\left( C P_{t-1|t-2} C^{T} + R \right)^{-1}
C P_{t-1|t-2} A^{T} \\[6pt]
&& +\, W
\end{array}
\right.
\end{equation}
Finally, the one-step-ahead prediction of the neural observations, $\hat{y}_{t+1|t}$, is computed by mapping $\hat{a}_{t+1|t}$ through the decoder defined in equation (\ref{methods-dfine-decoder}). Figure \ref{fig:DFINE-architecture}b illustrates the end-to-end process of using DFINE for future prediction. 

The forecasting horizon can be extended beyond one step by recursively applying the Kalman predictor on the latent dynamic  factors. Specifically, after computing the one-step-ahead prediction $\hat{x}_{t+1|t}$, the state is repeatedly propagated forward using the recurrence relation defined in equation (\ref{methods-dfine-lssm-xt}) (e.g., $\hat{x}_{t+K|t} = A^{K-1}\hat{x}_{t+1|t}$), yielding multi-step forecasts of the latent dynamics. The corresponding manifold factors can then be obtained as
\begin{equation}\label{methods:kalman-prediction-ksteps}
\hat{a}_{t+K|t} = C\hat{x}_{t+K|t}.
\end{equation}
Finally, the predicted manifold factors are mapped back to the observation space through the decoder $f_{\theta}$ to obtain the $K$-steps-ahead neural prediction $\hat{y}_{t+K|t}$.
In this way, DFINE performs long-horizon forecasting by rolling its linear dynamical model forward and decoding the resulting latent trajectory.

\subsubsection{Parameter optimization}\label{Methods: DFINE-parameter optimization}
DFINE is trained by optimizing a future-step-ahead prediction cost, leveraging the model's inference structure to align latent dynamics with observed neural activity. The optimization loss is defined as
\begin{equation} \label{methods-dfine-cost}
    L(\psi, \theta, \phi)=\frac{1}{K}\sum_{k=1}^{K}\frac{1}{T-k}\sum_{t=1}^{T-k} (y_{t+k}-\hat{y}_{t+k|t})^2,
\end{equation}
where $K$ is the maximum forecasting horizon (i.e., number of steps-ahead) into the future to predict. In our analyses, we train and evaluate models using $K=1$ (see Section \ref{results:longer-horizons} for an ablation on the value of $K$ used). The learnable parameters include $\psi=\{A, W, C, R, \mu_0, \Lambda_0\}$, which define the latent linear dynamical backbone as in equations (\ref{methods-dfine-lssm-xt}) and (\ref{methods-dfine-lssm-at}), and the parameters $\theta$ and $\phi$, which correspond to the encoder and decoder, respectively, as defined in equations (\ref{methods-dfine-encoder}) and (\ref{methods-dfine-decoder}). To train DFINE, we used Adam optimizer with an initial learning rate of $\eta=2\times 10^{-3}$ and trained for 400 epochs (see \ref{app:adam} for additional training details).

\subsection{Baselines}\label{methods-baselines}
\subsubsection{LSSM} \label{methods-lssm}

Latent linear state-space models with continuous Gaussian observations are formalized similar to the dynamic factors in DFINE (equations (\ref{methods-dfine-lssm-xt}) and (\ref{methods-dfine-lssm-at})), namely
\begin{equation}
\left\{
\begin{array}{cc}
    x_{t+1} = Ax_t+w_t\\
    y_t  = Cx_t+v_k
\end{array}
\right.
\end{equation}
where $x_t$, $w_t$, $A$, and $C$ are defined as before. The key difference is that the linear observations of LSSMs, that is $y_t$, correspond to the neural features directly, rather than the manifold latent factors, $a_t$. Notably, LSSMs can be recovered as a special case of DFINE, where the encoder and decoders are replaced by identity functions thus resulting in the manifold factors being the neural observations (i.e., $a_t=y_t$).

 The learnable parameters for LSSMs are $\psi=\left\{A, W, C, R, \mu_0, \Lambda_0\right\}$, matching that of DFINE for its linear dynamical backbone (Section \ref{Methods: DFINE-parameter optimization}). Here we use subspace system identification to estimate the model parameters\cite{vanoverscheeIntroductionMotivationGeometric1996}, as was done in prior works that modeled iEEG data using LSSMs \cite{yang_lssm_2019, ahmadipour2021_adaptive_tracking}. We used an implementation from a publicly available Python package (\href{https://github.com/ShanechiLab/PyPSID}{https://github.com/ShanechiLab/PyPSID}) to perform system identification. Finally, neural prediction using LSSMs can also be computed using the Kalman predictor with a formulation similar to equation (\ref{methods:kalman-prediction}).

\subsubsection{GRU}
The gated recurrent unit (GRU) \cite{gru_2014} is a recurrent neural network (RNN) architecture commonly used for time-series data due to its ability to capture nonlinear temporal dependencies~\cite{anumanchipalli2019_speech_synthesis, duraivelHighresolutionNeuralRecordings2023, nozari2024macroscopic, chungEmpiricalEvaluationGated2014, shenDeepLearningGated2018, xiaStackedGRURNNBasedApproach2021} and to regulate with its gating mechanism how past information is retained or overwritten by new inputs. Thus, we implemented a GRU-based forecasting model as our nonlinear dynamical model baseline. For a fair comparison against DFINE, we embedded the GRU between an encoder and decoder, such that the encoder mapped neural activity into a latent representation space, the GRU modeled the temporal dynamics in this latent space, and the decoder reconstructed the neural features from the outputs of the RNN backbone. Throughout this paper we refer to this GRU-based model as GRU.

We chose to implement the GRU using an encoder/decoder setup for two reasons. First, we wanted to ensure that DFINE did not have an unfair advantage due to increased modeling capacity. Second, it was important  that the only architectural difference between GRU and DFINE be the nonlinear vs. linear autoregressive backbones. This would allow us to more accurately evaluate DFINE's ability to capture nonlinear neural dynamics, i.e., if it performed on par with the GRU, while still enabling flexible inference with its linear autoregressive backbone. 

In our formulation, the GRU update equations are given by:
\begin{equation}\label{eq:gru}
\left\{
\begin{array}{lcl}
z_{t} &=& \sigma(W_z [x_{t-1}, a_t]) \\[6pt]
r_{t} &=& \sigma(W_r [x_{t-1}, a_t]) \\[6pt]
\tilde{x}_{t} &=& \tanh(W_x [r_t \odot x_{t-1}, a_t]) \\[6pt]
x_{t} &=& (1 - z_t) \odot x_{t-1} + z_t \odot \tilde{x}_{t},
\end{array}
\right.
\end{equation}
where $\odot$ represents the elementwise product, $a_t$ denotes the encoded input at time $t$ given by $a_t=f_\phi(y_t)$, $x_t$ is the dynamical latent state, and $z_t$ and $r_t$ are the update and reset gates in the GRU, respectively. For forecasting, a one-step-ahead prediction is obtained by mapping the latent state into the observation space as
\begin{equation} \label{methods-gru-forecasting}
\left\{
\begin{array}{lcl}
    \hat{a}_{t+1|t} &=& W_{out} x_t,\\
    \hat{y}_{t+1|t} &=& f_\theta(\hat{a}_{t+1|t})
\end{array}
\right.
\end{equation}
where $W_{out} \in \mathbb{R}^{n_x \times n_a}$ is a learnable matrix and $f_\theta$ denotes the decoder. Taken together, the parameters of the GRU model are $\psi=\left\{W_z, W_r, W_x, W_{out}, \theta, \phi \right\}$. We use the same training objective and optimizer settings as DFINE to fit the GRU parameters (see Section \ref{Methods: DFINE-parameter optimization} and \ref{app:adam}). Finally, similar to DFINE and LSSM, equation (\ref{methods-gru-forecasting}) can be extended to multi-step prediction by recursively feeding each predicted value, $\hat{y}$, back into the model $K$ times to obtain $\hat{a}_{t+K|t}$.

\subsubsection{Naive predictor}
As another baseline, we consider the naive one-step-ahead predictor defined as
\begin{equation}
    \hat{y}^{(naive)}_{t+1|t} = y_t,
\end{equation}
which simply uses the most recent observation as the next prediction.  
This naive predictor, commonly used in time-series forecasting, serves as a minimal benchmark to assess whether a proposed model provides predictive value beyond a trivial baseline \cite{yang_lssm_2019, hewamalageForecastEvaluationData2023}.

\subsection{Human iEEG data collection and processing}
\subsubsection{Human recordings}
We collected a new iEEG dataset from 10 epilepsy patients (S1-S10) implanted with intracranial electrodes for standard clinical monitoring unrelated to our research at the Keck Hospital of the University of the Southern California. Participants provided written informed consent to participate in our study. All experimental procedures and consent forms were approved by the University of Southern California’s Institutional Review Board (IRB). The neural signals were recorded using the XLTek EEG clinical recording system (Natus Medical, Inc.) with a sampling rate of 2048 Hz. The study utilized both strip and depth electrodes that covered a broad set of cortical and subcortical regions across subjects, including orbitofrontal, insular, temporal, and cingulate cortices, as well as the amygdala and hippocampus. 

\subsubsection{Preprocessing of raw iEEG signals}\label{methods-preproc}
Raw iEEG recordings were initially downsampled to 256 Hz using an order-8 IIR Chebyshev Type I anti-aliasing filter with a 100 Hz cutoff frequency. The downsampled signals were then visually inspected and segments contaminated by non-neural artifacts (e.g., motion-related noise) were excluded. Next, we applied a high-pass filter (order-2 IIR Butterworth, cutoff at 1 Hz) and a 60 Hz notch filter (order-4) to remove DC and line noise, respectively. Finally, signals were re-referenced using a common average across contacts within each electrode strip or depth probe. For each subject we used 140 hours of neural data, which was the shortest available recording length among all subjects.

\subsubsection{Power features} \label{methods-powerfeatures}
Spectral power band features have been shown to have physiological relevance and are commonly used to study diverse brain states \cite{sani2018_mood_decoding, shirvalkarFirstinhumanPredictionChronic2023, alagapanCingulateDynamicsTrack2023, beuterClosedloopCorticalNeuromodulation2014a, raoDirectElectricalStimulation2018, scangosPilotStudyIntracranial2020}. As such, here we chose to model the latent dynamics present in power band features computed from the preprocessed iEEG recordings (Section \ref{methods-preproc}).

To compute the power features, we first applied an order-8 Butterworth IIR filter to extract five frequency bands: 1–8 Hz (delta/theta), 8–12 Hz (alpha), 12–30 Hz (beta), 30–55 Hz (low gamma), and 65–100 Hz (high gamma). Then, for every contact and frequency band, the filtered signals were segmented into non-overlapping 10-second windows and the mean-squared amplitude was computed within each window. We then took the logarithm of these values to obtain log-power features. To ensure consistency across subjects, we randomly selected 40 contacts per patient, the minimum number of available contacts across all subjects, and used the log-power features of the selected contacts as the neural observations in our models (i.e., $n_y = 200$). In Appendix \ref{app:results-ss-seed012}, we performed an ablation on the contacts selected to ensure our results held across any set of randomly-selected contacts. 

\subsubsection{Cross-validation}
\begin{figure} 
    \includegraphics[width=1\linewidth]{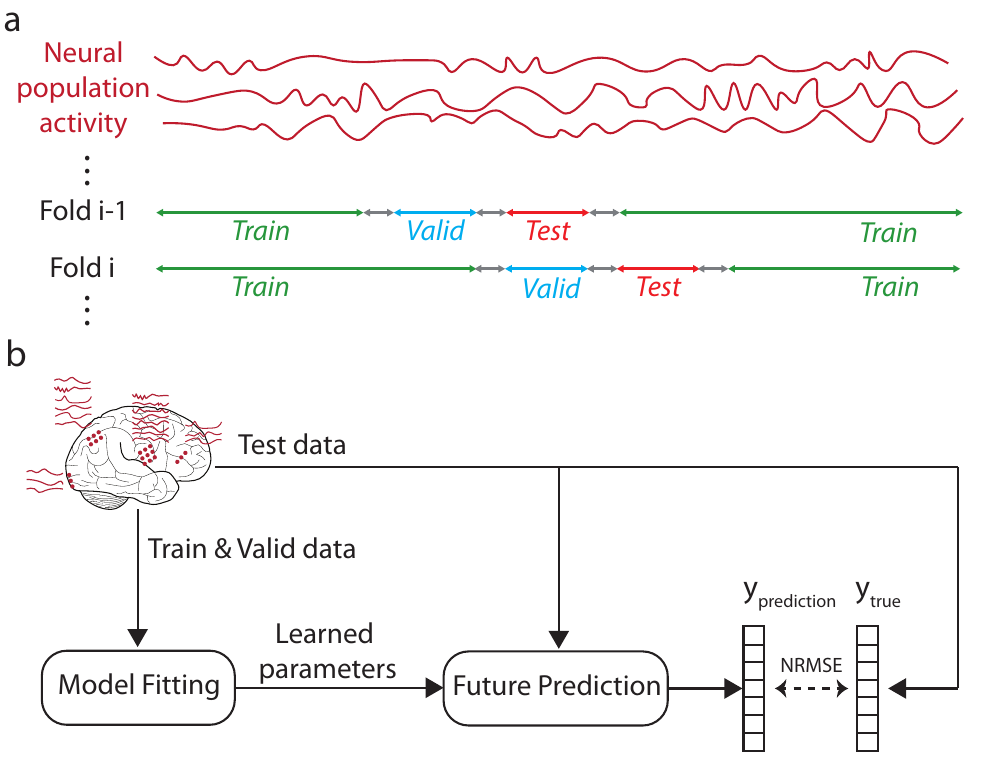}
    \caption{Experimental methodologies. (a) $N$-fold cross-validation of the continuous iEEG activity. The gray gaps are excluded to avoid information leakage across train, valid, and test sets. \if0Each gap window represents 1 hour.\fi (b) The model fitting and evaluation procedure for each of the $N$ folds.}
    \label{fig:methods-problem-setup} 
\end{figure}

We trained and tested our models using $N$-fold cross-validation. The continuous neural features were divided into $N$ folds, and in each iteration one fold was held out for testing while the preceding fold was used for validation. We z-score standardized the features using the mean and standard deviation computed from the training set. In our analysis, given the large amount of temporal data, we set $N=20$. We excluded the first fold from serving as a test set to ensure a consistent setup in which each test fold was always preceded by a validation fold. Further, to avoid potential information leakage between the training, validation, and test sets, three 1-hour buffer windows are excluded at each split boundary, as illustrated by the gray gaps in Figure \ref{fig:methods-problem-setup}.

\section{Results}\label{sec: Results}
\begin{figure*}[h]
    \includegraphics[width=1\linewidth]{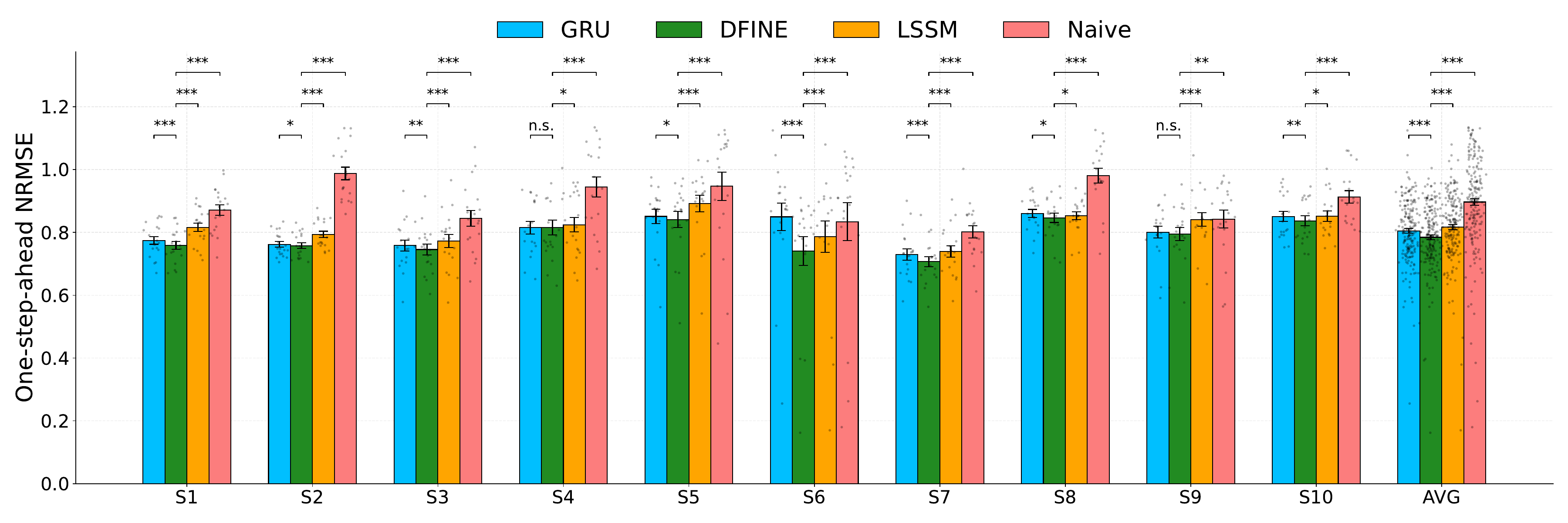}
    \caption{Across the population and within each subject, DFINE (green) outperforms LSSM (yellow) and the naive predictor (pink) on one-step-ahead neural prediction, while performing on par with (matching or exceeding) GRU (blue). Asterisks show significance (*: $p<0.05$, **: $p<0.01$, ***: $p<0.001$, and n.s.: $p>0.05$; one-sided and two-sided Wilcoxon with LSSM and GRU, respectively). Whiskers indicate s.e.m. and each dot represents the test performance for one held-out fold.}
    \label{fig:results-ss}
\end{figure*}

\subsection{DFINE outperforms LSSM in neural prediction}\label{results:lssm-vs-dfine}
\begin{figure}[h]
    \centering
    \includegraphics[width=1\linewidth]{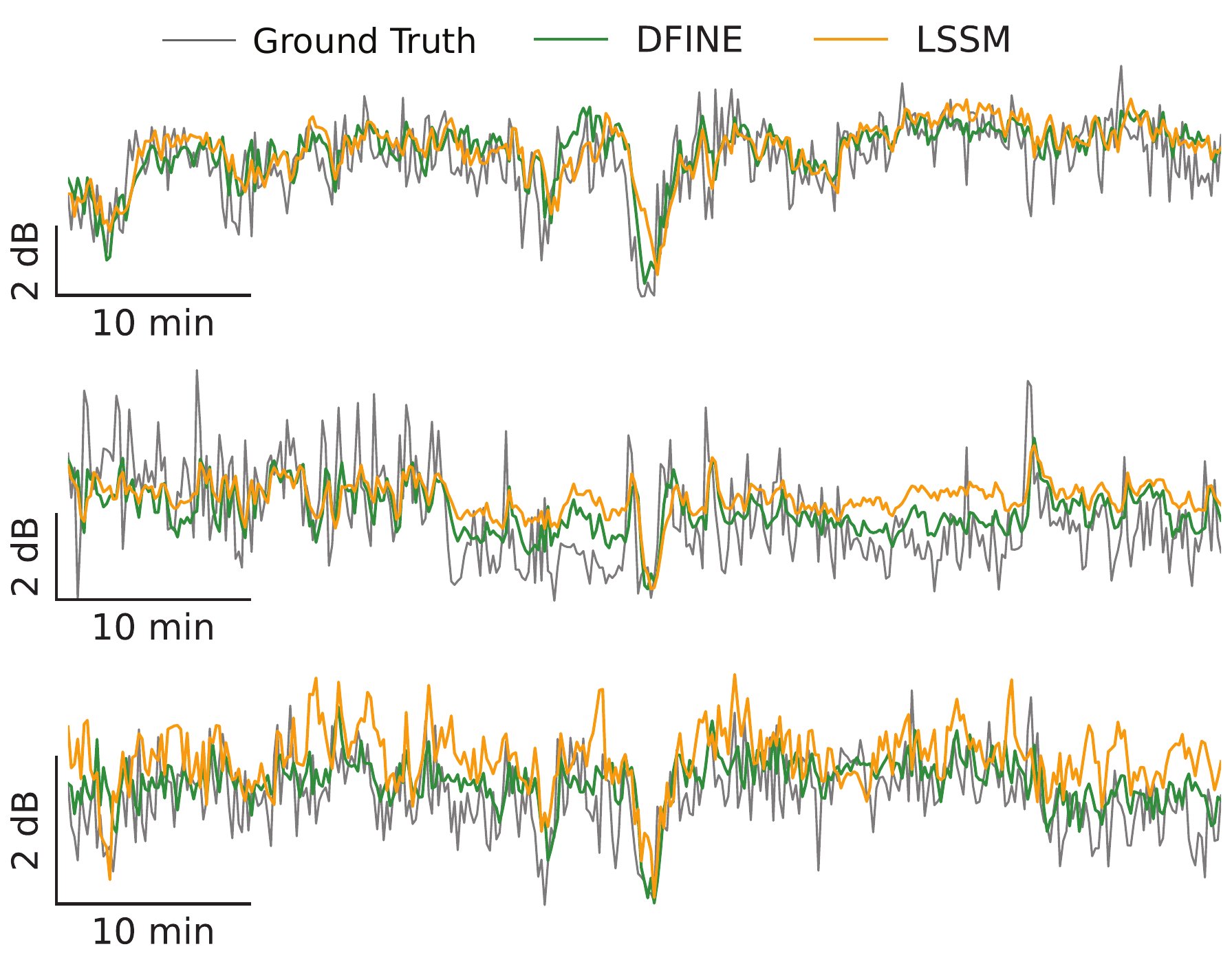}
    \caption{Three iEEG power features and their corresponding prediction traces by DFINE (green) and LSSM (yellow) in a single subject. DFINE clearly better predicts the future iEEG activity better than LSSM.}
    \label{fig:results-dfinevslssm-prediction}
\end{figure}

For our first comparison, we observe that DFINE significantly outperforms LSSMs in neural prediction across all 10 subjects. DFINE achieved a significantly smaller $NRMSE$ than LSSM both within each subject ($p<0.05$, one-sided Wilcoxon signed-rank test, $n=19$ samples per subject) and on the population average across all 10 subjects ($p<0.001$, one-sided Wilcoxon signed-rank test, $n=190$ samples pooled over all subjects). Population average and within subject $NRMSE$ results are presented in Figure \ref{fig:results-ss}. In Figure \ref{fig:results-dfinevslssm-prediction}, we present example prediction traces for a single subject, illustrating DFINE’s improved accuracy in capturing iEEG dynamics. 

Our results suggest that DFINE's ability to embed nonlinearity on top of a latent linear dynamical model can indeed help enhance the prediction of iEEG signals. By capturing nonlinear structure in the observations while retaining a linear state-space evolution, DFINE delivers improved predictive accuracy without sacrificing the desirable properties of linear dynamical models, such as flexible inference via Kalman filtering. Finally, it is important to note that DFINE achieves better neural prediction performance compared to the naive predictor, indicating that the model achieves a non-trivial prediction accuracy (see Figure \ref{fig:results-ss}).

\subsection{DFINE performs better than or similar to GRU in neural prediction while also enabling flexible inference}
We found that replacing the linear dynamical backbone with a nonlinear recurrent neural network did not improve the prediction of neural features. Indeed, in 8 of the 10 subjects DFINE performed significantly better than GRU ($p<0.05$, two-sided Wilcoxon signed-ranked test, $n=19$ samples per subject), while in the remaining 2 subjects (S4 and S9) the models performed comparably with no significant difference ($p>0.05$, two-sided Wilcoxon signed-rank test, $n=19$ samples per subject). Notably, DFINE outperforms GRU in neural prediction on the population average across all 10 subjects ($p<0.001$, two-sided Wilcoxon signed-rank test, $n=190$ samples pooled over all subjects). Our one-step-ahead $NRMSE$ results are presented in Figure \ref{fig:results-ss}.

Taken together, this evidence suggests that using a linear dynamical backbone does not impair DFINE's ability to accurately model nonlinearities in neural activity, while still allowing for flexible inference -- a capability not supported by RNNs. 

\subsection{DFINE stays superior in predicting longer horizons} \label{results:longer-horizons}
\begin{figure}
    \centering
    \includegraphics[width=1\linewidth]{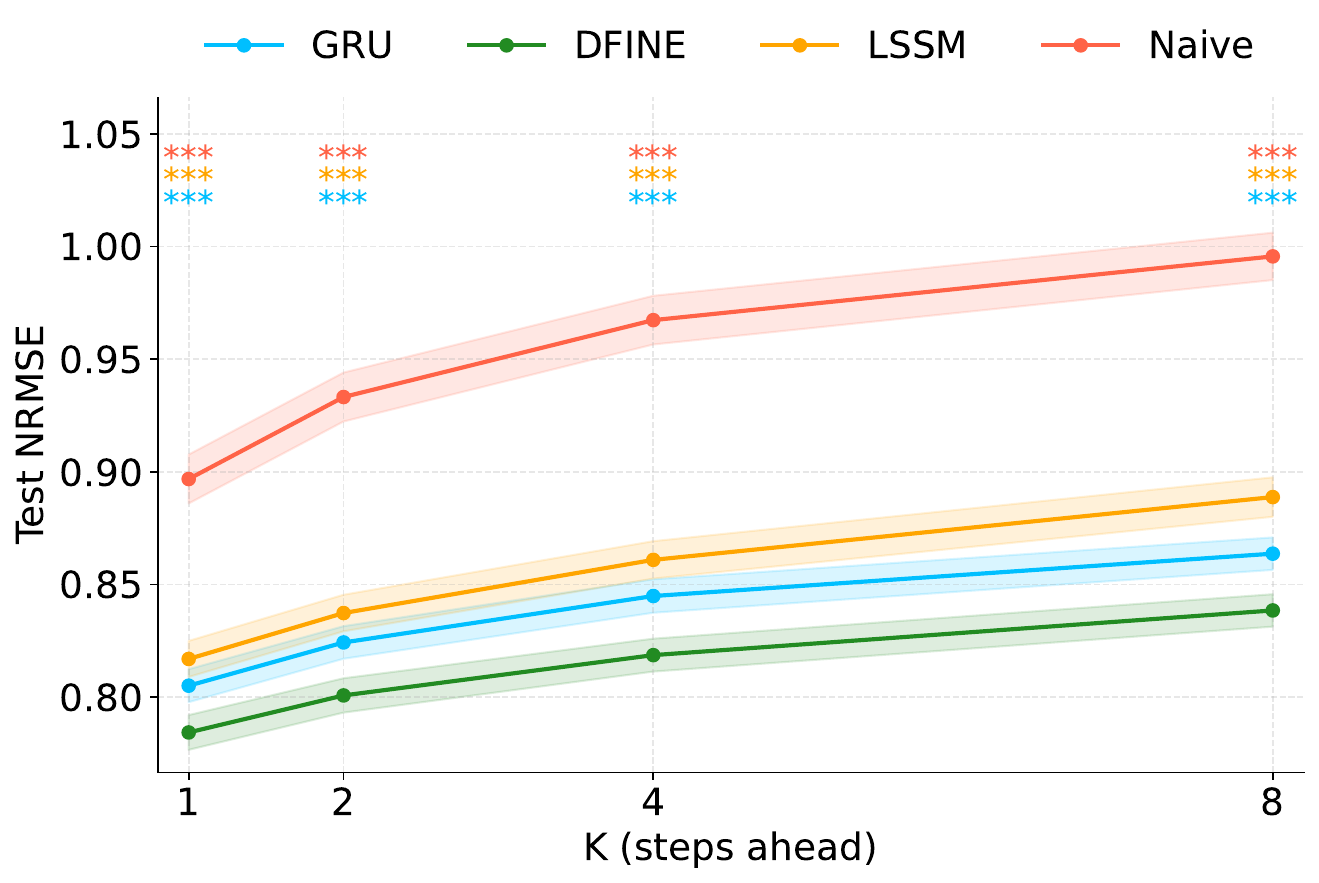}
    \caption{DFINE maintains superior multi-step-ahead prediction accuracy, outperforming all baselines. Solid curves represent the mean and the shaded regions denote the standard error of the mean (SEM) of all test folds pooled across the 10 subjects. Asterisks indicate significance (***:$p<0.001$, blue, yellow, and red indicating comparisons against GRU, LSSM, and naive, respectively).}
    \label{fig:results-DFINE-ksteps}
\end{figure}

Although one-step-ahead prediction serves as our primary evaluation metric, we further demonstrate that DFINE maintains its advantage when forecasting multiple steps into the future. For this, we retrained DFINE and GRU on adjusted losses using $K\in\{1,  2, 4, 8\}$; we then evaluated neural prediction accuracy for each of these horizon values, $K$. As shown in Figure \ref{fig:results-DFINE-ksteps}, DFINE consistently outperforms all baseline models across every tested prediction horizon ($p<0.001$, one-sided Wilcoxon signed-rank test, $n=190$ pooled samples). Importantly, both DFINE and the GRU baseline were trained and evaluated under the same multi-step forecasting horizons, ensuring a fair comparison.

DFINE’s superior performance across different horizon values demonstrates the model's ability to accurately capture the underlying temporal dynamics thereby achieving better long-term prediction results compared to the baselines.

\subsection{DFINE achieves flexible inference with missing observation}
We next evaluated DFINE's ability to perform flexible inference when portions of the iEEG data are missing, and found that DFINE was indeed able to achieve more accurate neural prediction, as compared to the baseline models. To simulate missing-data conditions, a common occurrence in wireless BCI settings, we randomly removed multiple continuous 40-second blocks (4 samples) of neural data. We then assessed one-step-ahead prediction accuracy across various observation ratios, where observation ratio is defined as $1-\frac{\mathrm{\# dropped\ samples}}{\mathrm{\# total\ samples}}$. For this analysis, we trained fresh models of DFINE and GRU using a forecasting horizon of $K=4$ (see equation \ref{methods-dfine-cost}). By doing so, we aimed to provide models with a broader temporal context and thus more fairly evaluate their ability to handle missing observations.

\begin{figure} 
    \centering
    \includegraphics[width=1\linewidth]{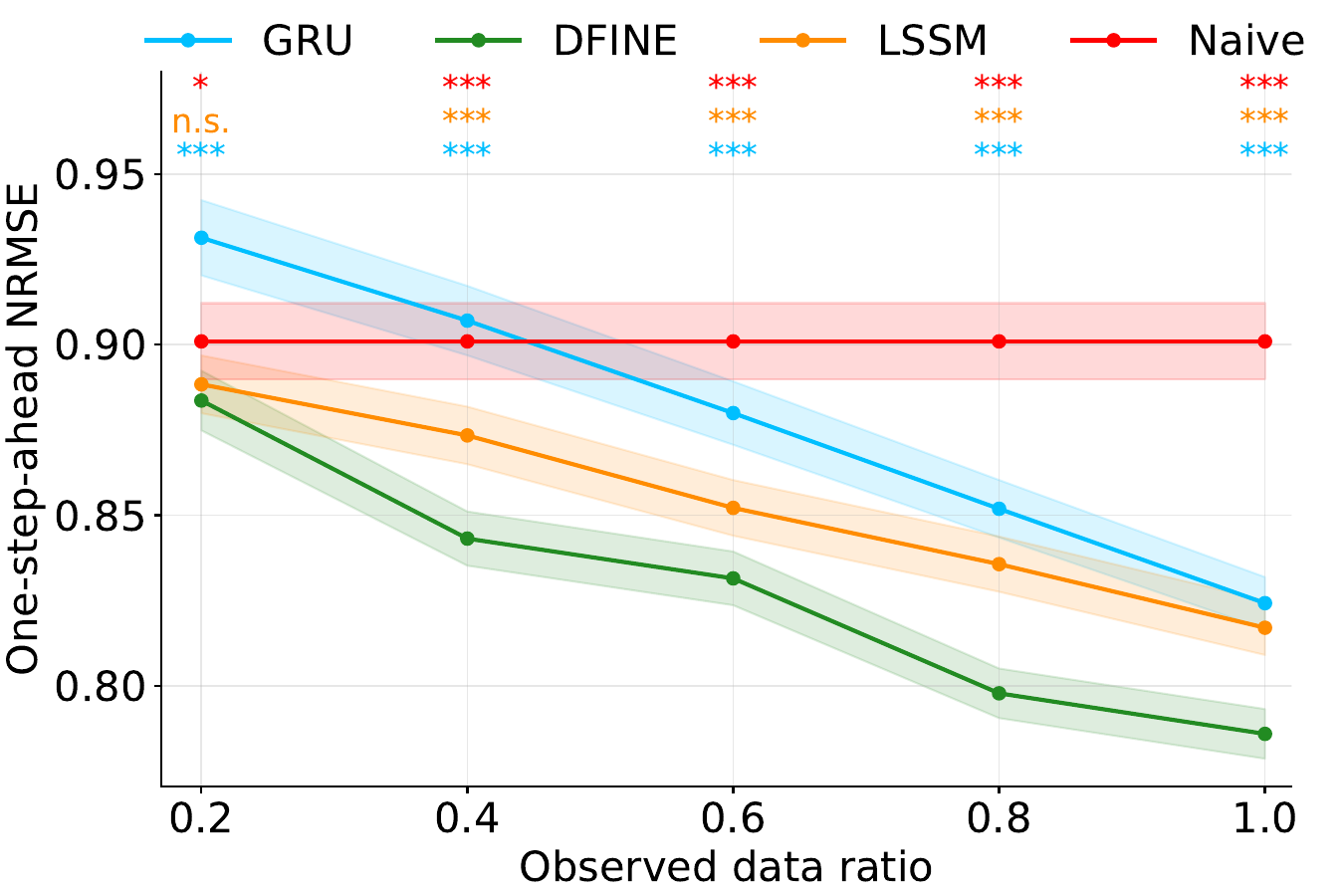}
    \caption{DFINE consistently achieved lower one-step-ahead neural prediction error compared to baseline models across most observation ratios. Multiple continuous blocks of neural data were randomly removed to simulate different observation ratios. DFINE consistently achieved lower prediction error than GRU across all observation ratios, and outperformed LSSM except at the lowest ratio (20\%), indicating improved robustness to missing neural samples. Solid curves represent the mean and the shaded regions denote the standard error of the mean (SEM) of all test folds pooled from 10 subjects. Asterisks are as defined in Figure \ref{fig:results-DFINE-ksteps}}
    \label{fig:results-missing-values}
\end{figure}

Across all observation ratios tested, DFINE achieved higher prediction accuracy than GRU, indicating that the model can more flexibly compensate for missing samples (Figure~\ref{fig:results-missing-values}, $p<0.001$, one-sided Wilcoxon signed-rank test, $n=190$ samples pooled over all subjects).
Indeed, DFINE is able to do so by performing Kalman filtering on the latent linear dynamics, thereby allowing the model to naturally handle periods with no observations by updating its estimate of the current latent state using only the recurrent dynamics (i.e., setting the Kalman gain equal to zero in equation \ref{methods:kalman-prediction}). In contrast, GRU requires an input at \emph{every} time step, indicating that the model's update formulation does not implicitly handle missing observations In our analysis we emulated missing samples by passing in in zero-imputed values to the model -- as is common practice for RNNs operating on incomplete data \cite{liptonDirectlyModelingMissing2016, erturkDynamicalModelingNonlinear2025}. Furthermore, we also found that DFINE outperformed LSSM across all observation ratios ($p<0.001$, one-sided Wilcoxon signed-rank test, $n=190$ samples pooled over all subjects), except at the 20\% level where their performance was comparable (i.e., no significant difference).

Taken together, these results demonstrate DFINE’s ability to more robustly handle missing iEEG measurements compared to baseline models, which is an important capability for real-world BCI settings wherein signal interruptions are common.

\begin{figure*}[h!]
    \centering
    \includegraphics[width=1\linewidth]{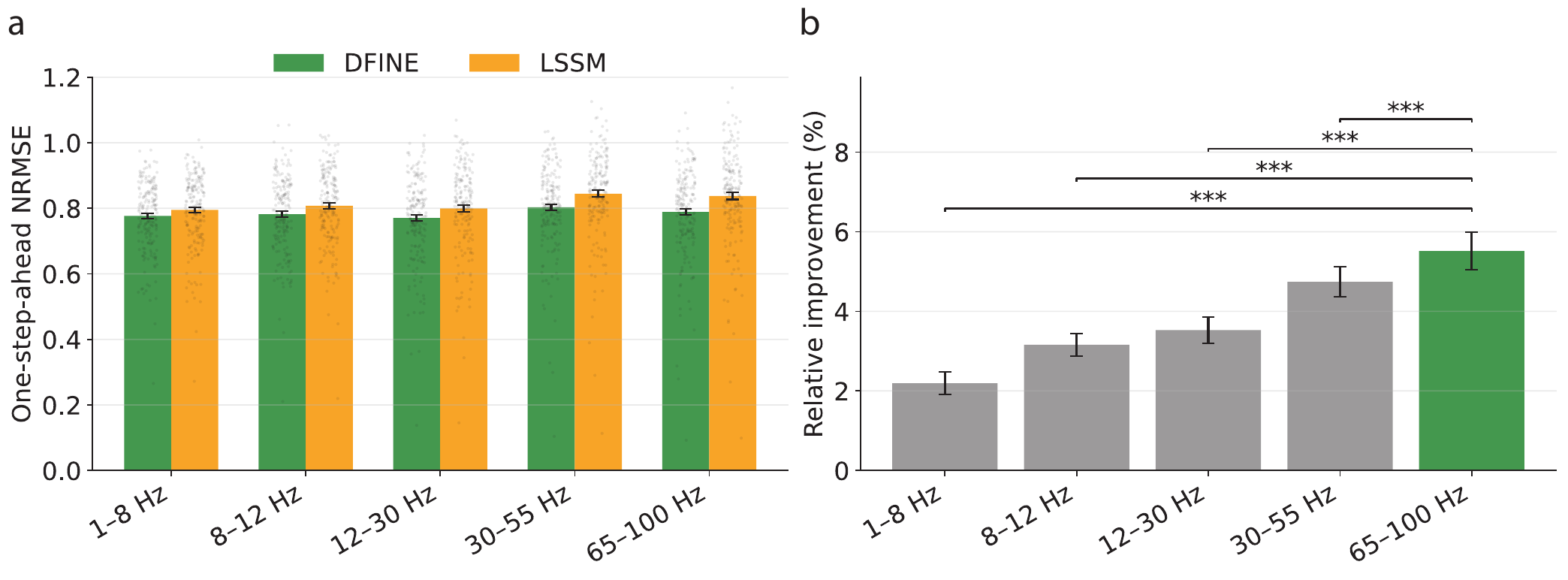}
    \caption{DFINE's model architecture yields the greatest improvement over LSSM in neural prediction when modeling 65-100 Hz power band activity. (a) One-step-ahead prediction accuracy of DFINE and LSSM per band pooled over all subjects. Bars represent the average NRMSE values and whiskers show standard error of measure (SEM), with each dot denoting a test value from a held-out fold. (b) High gamma (65-100 Hz) exhibited significantly greater relative improvement compared to the other power bands. Bars and whiskers are defined as in panel (a). Asterisks show significance (***: $p<0.001$, one-sided Wilcoxon signed-rank test, $n=190$ samples pooled over all subjects).}
    \label{fig:results-bandwise}
\end{figure*}
\subsection{DFINE offers the greatest improvement over LSSM when modeling high frequency power band features}
The results in Section \ref{results:lssm-vs-dfine} demonstrated that DFINE improved neural prediction accuracy compared to LSSM. To determine which frequency components benefited the most from DFINE's model architecture , we analyzed prediction accuracy for each of the 5 spectral bands considered here (see Section \ref{methods-powerfeatures}).
To do so, we compared the prediction accuracy of DFINE and LSSM within each frequency band. First, consistent with prior iEEG dynamical modeling work \cite{yang_lssm_2019}, we observed that LSSM achieved lower prediction error in the lower-frequency bands (1-8, 8-12, and 12-30~Hz) compared to the higher-frequency bands (30-55 and 65-100~Hz); results are presented in Figure \ref{fig:results-bandwise}a. We then evaluated DFINE's performance in each power band and quantified the relative improvement in prediction accuracy over LSSM. Relative improvement for band $b$ was computed as
\begin{equation}
    \mathrm{\Delta}_{b}
    =
    \frac{\mathrm{NRMSE}_{\mathrm{LSSM},\, b} - \mathrm{NRMSE}_{\mathrm{DFINE},\, b}}
         {\mathrm{NRMSE}_{\mathrm{LSSM},\, b}}
    \times 100\%,
\end{equation}
where $\Delta$ represents the relative improvement in prediction accuracy and $b$ indexes one of the five frequency bands, 1-8~Hz, 8-12~Hz, 12-30~Hz, 30-55~Hz, and 65-100~Hz. While DFINE improved prediction accuracy across all five spectral bands, the largest relative gain was observed in the high frequency range (65-100~Hz), corresponding to high gamma activity. This improvement was significantly greater than that observed in the other four bands ($p<0.001$, one-sided Wilcoxon signed-rank test, $n = 190$ samples pooled across subjects). DFINE prediction accuracy results are presented in Figure \ref{fig:results-bandwise}a and relative improvement numbers are presented in Figure \ref{fig:results-bandwise}b.

Because high gamma power is widely used in both neuroscience and clinical applications, improved modeling in this band is particularly valuable.
 (See Discussion \ref{discussion: gamma high})

\section{Discussion}\label{sec: Discussion}
Here we demonstrated effective dynamical modeling of multisite intracranial human brain activity while also enabling flexible inference of latent states. We did so by investigating the extension of DFINE to modeling these multisite recordings.\emph{First}, we showed that DFINE outperforms linear state-space models (LSSMs) in neural prediction of iEEG data. \emph{Second}, we found that DFINE performs better than or similar to a GRU-based RNN model, while enabling flexible inference unlike this RNN model. This result suggests that DFINE's linear dynamical backbone when paired with a jointly learned nonlinear embedding may be sufficiently expressive for modeling the dynamics of iEEG data. \emph{Third}, we evaluated each model's future predictive accuracy (i.e., forecasting) for longer horizons and found that DFINE outperforms all baselines across the population of subjects. \emph{Fourth}, we tested each model's robustness to missing neural samples, observing that DFINE's model architecture enables improved flexible inference in the presence of missing iEEG samples. \emph{Fifth}, we examined DFINE’s relative improvement across spectral bands and found that its largest gains over LSSM occur in the higher-frequency (high gamma) features. Taken together, these results position DFINE as an effective, flexible dynamical modeling framework for capturing nonlinear dynamics in human iEEG, enabling accurate characterization of large-scale brain activity and supporting the development of next-generation BCIs.

\subsection{Significance of modeling iEEG data}
Intracranial EEG (iEEG) provides a good combination of resolution, coverage, and recording stability, which makes it a useful modality for studying large-scale, multisite neural dynamics and for developing  neurotechnologies. Unlike noninvasive scalp EEG, iEEG sensors are positioned directly on or within brain tissue. This can enable  higher signal-to-noise ratios and reduce the blurring effects introduced by the skull and scalp \cite{buzsaki2012_extracellular_fields}. Furthermore, iEEG supports continuous stable recordings over extended time periods, making it an attractive modality for studying behavioral and internal states that evolve over long time-periods and for building stable neural decoders \cite{shanechiBrainMachineInterfaces2019a, ahmadipour2021_adaptive_tracking, llo_reviewpaper_2024}. Additionally, iEEG provides broad spatial coverage, often spanning multiple cortical and subcortical regions, allowing for the study of network-level large-scale interactions in the human brain.

\subsection{Choice of latent state dynamical modeling}
In this work, we focused on latent state dynamical modeling given its demonstrated utility for neuroscientific studies and BCI  applications. Such models provide a powerful framework to capture the spatiotemporal patterns in neural data with low-dimensional latent states that aggregate information over time. Furthermore, they have demonstrated utility for decoding various behavioral and internal states from iEEG, for example decoding mood \cite{sani2018_mood_decoding}, cognitive control \cite{basuClosedloopEnhancementNeural2023}, chronic pain \cite{shirvalkarFirstinhumanPredictionChronic2023}, and speech\cite{anumanchipalli2019_speech_synthesis, duraivelHighresolutionNeuralRecordings2023}.  Latent state dynamical models also have great potential for designing model-based closed-loop BCIs such as closed-loop neuromodulation systems\cite{shanechiBrainMachineInterfaces2019a,llo_reviewpaper_2024,hoang2017biomarkers, yangControltheoreticSystemIdentification2018, yang2021modelling, sani2021_closed_loop_bci ,bolusStatespaceOptimalFeedback2021}. They have also had great success in developing BCIs for restoration of motor function \cite{Gilja2012HighPerformance,Sussillo_2012, orsborn2012closed, shanechi2016robust, shanechi2017rapid, nason2020low, Nason2021Realtime} and for studying inter-regional neural dynamics \cite{jha2025prioritized}. Here we demonstrated  nonlinear and flexible latent state dynamical modeling of human iEEG recordings. Our model outperformed prior dynamical latent state models of iEEG, including LSSMs and RNNs, thus showing utility for neuroscience and neurotechnology. 

\subsection{Nonlinear modeling of iEEG dynamics with flexible inference}
A latent state dynamical model for robust BCIs should not only enable accurate nonlinear modeling but also achieve flexible inference in regimes with missing samples. LSSMs have been previously used to effectively model iEEG data \cite{sani2018_mood_decoding, yang_lssm_2019, ahmadipour2021_adaptive_tracking, shirvalkarFirstinhumanPredictionChronic2023}. While LSSMs satisfy flexible inference, they do not model nonlinearity. Consistent with this, we found that DFINE outperformed LSSMs in neural prediction. 

We also compared to RNNs, which are commonly used in neuroscience and for modeling nonlinearity in iEEG \cite{anumanchipalli2019_speech_synthesis, duraivelHighresolutionNeuralRecordings2023, nozari2024macroscopic}, but do not address flexible inference. In particular, we compared DFINE with a GRU-based recurrent model, a widely-used, because it is a modern and effective RNN for for modeling time-series data \cite{gru_2014, chungEmpiricalEvaluationGated2014, shenDeepLearningGated2018, xiaStackedGRURNNBasedApproach2021}. Across 10 subjects, DFINE matched or exceeded GRU performance, indicating that a linear dynamical backbone can be sufficiently expressive when paired with a jointly learned nonlinear embedding, while having the additional advantage of flexible inference. We further found that DFINE’s benefits extend to predictions beyond the immediate next step. Indeed, DFINE outperformed all baselines across longer forecasting horizons.

Moreover, we evaluated DFINE’s ability to perform flexible inference in the presence of missing iEEG segments, a common challenge in real-world wireless BCIs. DFINE achieved higher prediction accuracy than GRU across all observation ratios. This robustness reflects another benefit of using a linear dynamical backbone, which allows the use of Kalman filtering to directly account for missing observations.

Taken together, these results demonstrate the ability for nonlinear and flexible modeling of multisite human iEEG activity, even in the presence of missing neural samples. These properties address key requirements for advancing BCI and closed-loop neuromodulation systems.

\subsection{Improvement in modeling of high frequency spectral features} \label{discussion: gamma high}
Here we focused on modeling spectral power band features derived from iEEG. Spectral power has been extensively used in neuroscience, for example to identify neural biomarkers across various disorders \cite{littleFunctionalRoleBeta2014,gadhoumiSeizurePredictionTherapeutic2016,raoDirectElectricalStimulation2018, scangosPilotStudyIntracranial2020,scangosClosedloopNeuromodulationIndividual2021, shirvalkarFirstinhumanPredictionChronic2023}. Furthermore, various studies have demonstrated that power across canonical frequency bands reflects fundamental aspects of cognitive and affective processing \cite{sani2018_mood_decoding, jacobsHippocampalThetaOscillations2014, johnsonIntracranialRecordingsHuman2015, buzsaki2012_extracellular_fields, pesaranInvestigatingLargescaleBrain2018}. These properties together make spectral power features a meaningful target for dynamical modeling. Interestingly, we found that DFINE achieved its largest relative gains compared to LSSM in the high frequency (high gamma) range, potentially indicating a higher degree of nonlinearity in these features. Prior studies have shown that high gamma activity carries information about specific cortical computations, such as movement-related activity in motor cortex \cite{gunduzDifferentialRolesHigh2016}, and speech production and perception in language areas \cite{bhaya-grossmanSharedLanguagespecificPhonological2025}. Clinically, high gamma power has also been used for functional mapping during neurosurgical planning \cite{croneFunctionalMappingHuman1998} and for identifying seizure onset zones \cite{jacobsHighfrequencyOscillationsHFOs2012}. Thus, improved modeling of high gamma activity may be particularly relevant for a variety of downstream applications.

\subsection{Future directions}
Several important extensions remain for future investigation. First, future work may investigate multimodal models that integrate iEEG with complementary neural or physiological measurements. Prior works have shown the benefit of multimodal neural modeling \cite{stavisky2015high,hsiehMultiscaleModelingDecoding2018, abbaspourazadMultiscaleDynamicalModeling2019, wang2019estimating, abbaspourazadMultiscaleLowdimensionalMotor2021a, lu2021multi,durstewitz_multi_teacher_22, durstewitz_mmplrnn_22 ,ahmadipourMultimodalSubspaceIdentification2024, erturkDynamicalModelingNonlinear2025, erturkCrossModalRepresentationalKnowledge2025}, for example for fusing spiking activity with LFP measurements recorded with intracortical arrays in animal models \cite{stavisky2015high,hsiehMultiscaleModelingDecoding2018, abbaspourazadMultiscaleDynamicalModeling2019, abbaspourazadMultiscaleLowdimensionalMotor2021a, ahmadipourMultimodalSubspaceIdentification2024, erturkDynamicalModelingNonlinear2025, erturkCrossModalRepresentationalKnowledge2025}. Indeed, a recent work developed nonlinear multimodal dynamical models of intracortical spiking-LFP modalities with the ability for flexible inference \cite{erturkDynamicalModelingNonlinear2025} and demonstrated the model's utility in non-human primate data. As such, opportunities for multimodal iEEG modeling exist in settings where iEEG is collected alongside other physiological or neural modalities, such as fMRI\cite{berezutskayaOpenMultimodalIEEGfMRI2022}.

Furthermore, here we focused on unsupervised modeling of multisite human brain recordings. Another future direction is to explore the joint modeling of behavior and iEEG activity. Indeed, prior works have introduced dynamical models for joint neural-behavioral data that dissociate behaviorally relevant latent states \cite{saniModelingBehaviorallyRelevant2021, sani_dissociative_2024, oganesianSpectralLearningShared2024, hosseini2025dynamical,hurwitz_tndm_21} by utilizing linear, generalized linear, or RNN dynamical model architectures. Also, DFINE has been extended to a supervised setting with a mixed neural-behavioral loss \cite{dfine_2024}. These methods, however, have so far largely been developed and evaluated on intracortical recordings in non-human primates. Studying whether and how multisite human iEEG can be jointly modeled with behavioral data while enabling flexible inference is an interesting future direction. 

Another important direction for future work is to allow for the incorporation of inputs, such as sensory stimuli or neuromodulation inputs, when modeling iEEG. Prior works have developed linear input-driven dynamical models of intracortical activity with neurostimulation input \cite{yang2021modelling}. Furthermore, input-driven latent state dynamical models have been developed for neural-behavioral data, using both linear \cite{vahidi2024} and nonlinear \cite{vahidi2025braid} RNN dynamical architectures. As such, explicitly modeling the influence of external inputs in our models of iEEG is an important direction that can lead to closed-loop BCI systems and provide insight into intrinsic vs. input-driven human brain network dynamics. One other interesting direction is to study the geometry \cite{chaudhuri2019intrinsic,hsieh2025probabilistic,dfine_2024} of the latent factors that DFINE recovers in iEEG dynamics across different behavioral tasks and contexts.

Finally, a future challenge is to address non-stationary brain dynamics. Some prior approaches have relied on adaptive linear filtering techniques \cite{ahmadipour2021_adaptive_tracking, yang2016adaptive, yang2019developing, yang2021adaptive, hsieh2018optimizing,shanechi2016robust} or switching linear dynamical models to track non-stationarity \cite{linderman_bayesian_2017,huModelingLatentNeural2024, songModelingInferenceMethods2022, songUnsupervisedLearningStationary2023}. Other methods have also been developed, for example using batch-based updating of model parameters for spiking activity \cite{orsborn2012closed,Gilja2012HighPerformance} or using alignment of latent subspaces in spiking activity to enable stability over time \cite{degenhart2020stabilization,karpowicz2025stabilizing, farshchian2018adversarial,bishop2014self}. However, nonlinear adaptive modeling of iEEG remains unexplored and is another important future direction that will be especially important for maintaining stable performance over time in BCIs.

\section*{Acknowledgments}
The authors acknowledge support of the National Institutes of Health (NIH) grants R01MH123770, R61MH135407, and RF1DA056402, the One Mind Rising Star Award, and the Foundation for OCD Research. We would like to thank Dr. Danil Tyulmankov and Eray Erturk in the NSEIP Lab for their valuable feedback.
\section*{References}
\bibliographystyle{ieeetr}
\bibliography{references_1, references_2}
\newpage
\clearpage
\onecolumn
\appendix              

\section*{Appendices}

\setcounter{table}{0}
\renewcommand{\thetable}{S\arabic{table}}%

\section{Training details}
\label{app:adam}

In this work, we used the Adam optimizer with a learning rate of $\eta = 2 \times 10^{-3}$. A step-based learning rate scheduler was applied, with step size $s = 100$ epochs and decay factor $\gamma = 0.5$, for a total of $400$ training epochs. During training, we monitor the validation loss at each epoch and select the model checkpoint that achieves the lowest validation loss. The same optimization settings were applied for training both DFINE and GRU.

\section{Encoder and decoder architectural details}
\label{app:encdec}

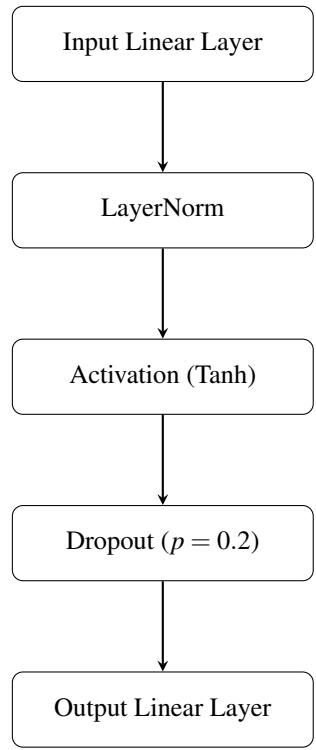
\begin{figure}[H]
\centering
\begin{tikzpicture}[
    layer/.style={draw, rectangle, rounded corners, minimum width=4cm, minimum height=1cm, align=center},
    arrow/.style={-stealth, thick},
    node distance=1.2cm
]

\node[layer] (linear1) {Input Linear Layer};
\node[layer, below=of linear1] (norm1) {LayerNorm};
\node[layer, below=of norm1] (act1) {Activation (Tanh)};
\node[layer, below=of act1] (drop1) {Dropout ($p=0.2$)};
\node[layer, below=of drop1] (output) {Output Linear Layer};

\draw[arrow] (linear1) -- (norm1);
\draw[arrow] (norm1) -- (act1);
\draw[arrow] (act1) -- (drop1);
\draw[arrow] (drop1) -- (output);

\end{tikzpicture}
\caption{
{\bf Architecture of the single hidden-layer MLP used in DFINE and GRU for both the encoder and decoder networks.}
Each block consists of a linear projection, layer normalization, Tanh activation, dropout, and a final output linear projection layer.
}
\label{app:mlp-blocks}
\end{figure}

\clearpage

\section{Additional figures}
\label{app:extra}

\begin{figure}[H]
\centering
\includegraphics[width=0.95\linewidth]{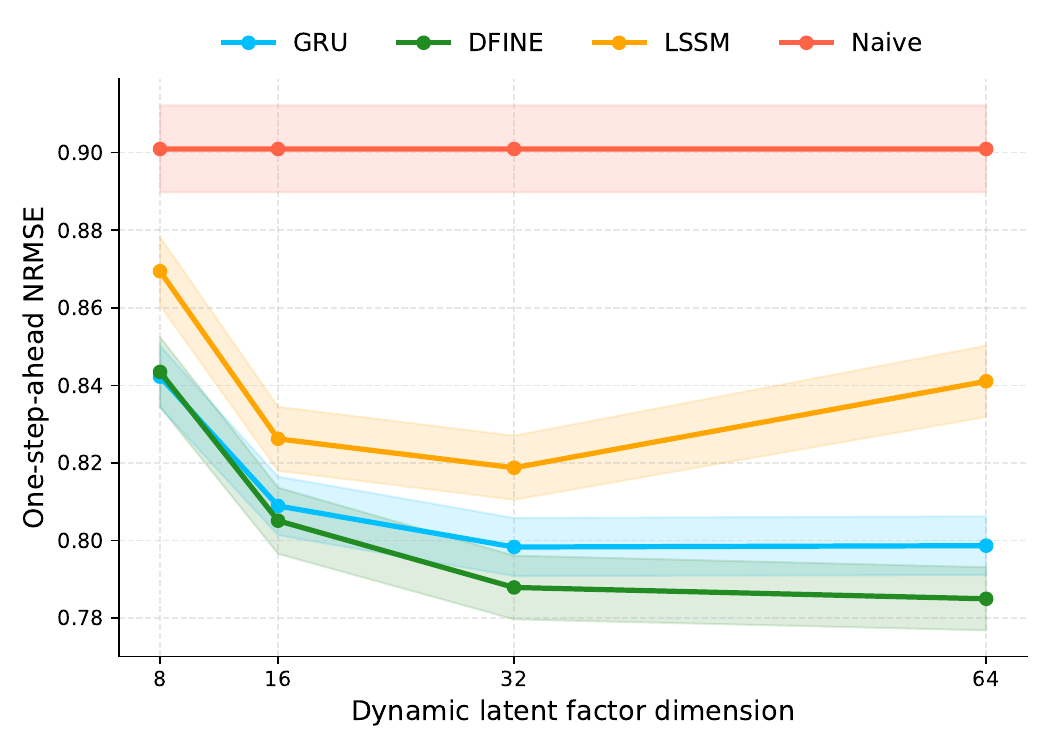}
\caption{{\bf Neural prediction accuracy on the validation folds as a function of the latent dynamic factor dimensionality, $n_x$.} Solid curves represent the mean and the shaded regions denote the standard error of the mean (SEM) of all \emph{validation} folds pooled across 10 subjects. Here we chose to evaluate on the validation and not the test folds because $n_x$ is treated as a hyperparameter in our framework. }
\label{app:nx}
\end{figure}

\begin{figure}[H]
\centering
\includegraphics[width=0.475\linewidth]{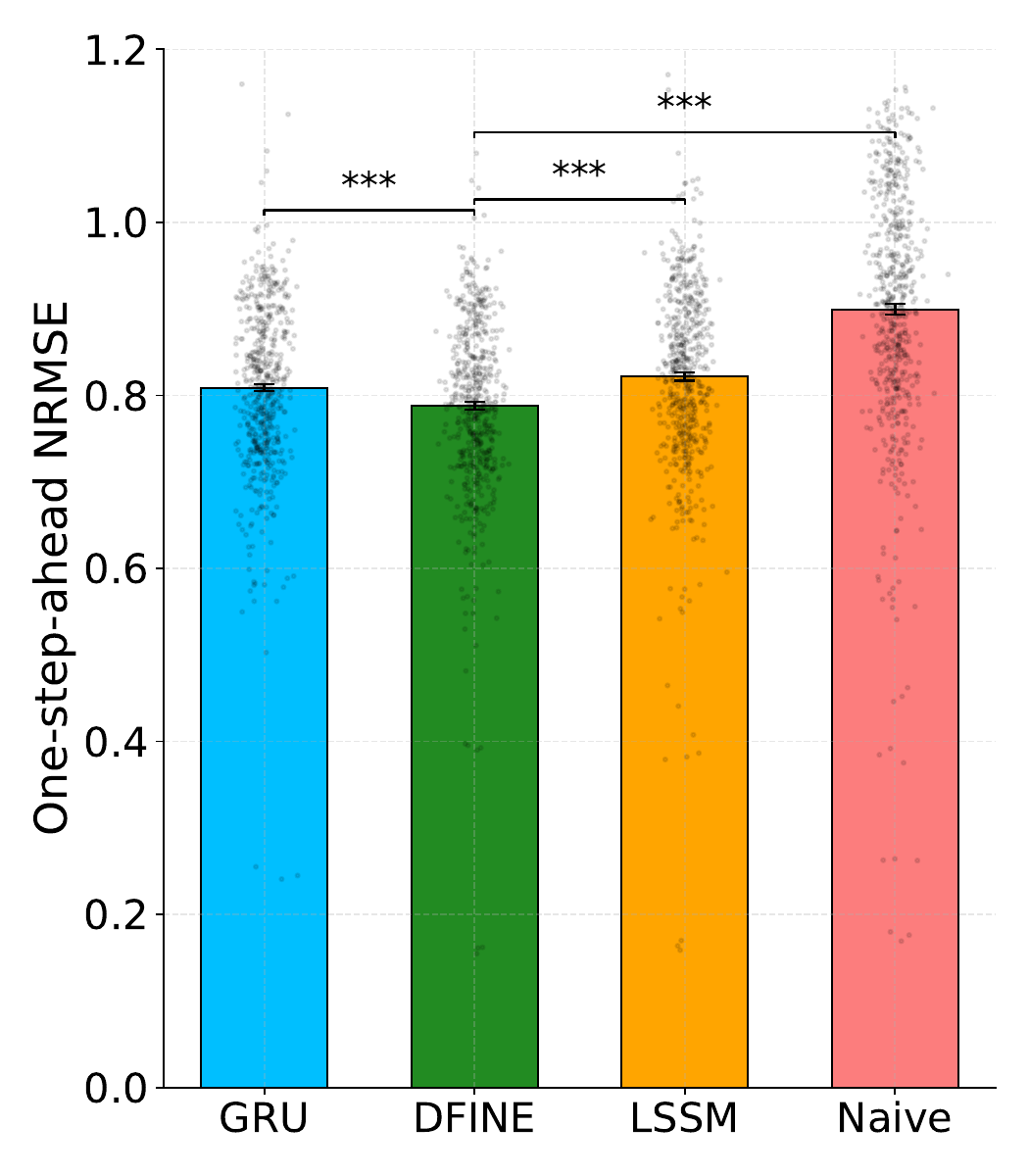}
\caption{
{\bf DFINE achieved higher neural prediction accuracy compared to baseline models across 3 different randomly-selected channel subsets per subject.}
For each of the 3 seeds used, 40 contacts were randomly sampled to construct the
log-power feature set ($n_y = 200$) used during training and evaluation. Bars show mean NRMSE across subjects and seeds, with scatter points
indicating individual fold-level values. The consistent relative ordering
across all seeds demonstrates that DFINE’s performance advantage is not
specific to a particular channel selection ($p<0.001$, one-sided Wilcoxon signed-rank test, $n=570$ pooled across all subjects and seeds).}
\label{app:results-ss-seed012}
\end{figure}

\end{document}